\documentclass[reprint,superscriptaddress,amsmath,amssymb,aps,pra]{revtex4-2}

\usepackage{comment}
\usepackage{enumerate}
\usepackage{graphicx}
\usepackage{dsfont}
\usepackage{dcolumn}
\usepackage{bm}
\usepackage{xcolor}
\definecolor{grape}{RGB}{63, 37, 110}
\usepackage{hyperref}
\hypersetup{
    colorlinks=true,
    linkcolor=grape,
    filecolor=grape,      
    urlcolor=grape,
    citecolor=grape
    }
\usepackage{tikz}
\usetikzlibrary{quantikz}
\usepackage{dirtytalk}
\usepackage{float}

\newcommand{\dd}{\text{d}}
\newcommand{\g}{\text{g}}
\newcommand{\lr}{\rangle \langle}

\newcommand{\tr}{\text{Tr}}
\def\id{\mathds{1}} 
\newcommand{\figref}[1]{Fig.~\ref{#1}}
\newcommand{\appref}[1]{App.~\ref{#1}}
\newcommand{\secref}[1]{Sec.~\ref{#1}}

\begin{document}

\preprint{APS/123-QED}
\title{Benchmarks for quantum communication via gravity}
\author{Kristian Toccacelo}
\email{kristo@dtu.dk}
\author{Ulrik Lund Andersen}
\email{ulrik.andersen@fysik.dtu.dk}
\author{Jonatan Bohr Brask}
\email{jonatan.brask@fysik.dtu.dk}

\affiliation{Center for Macroscopic Quantum States (bigQ), Department of Physics,
Technical University of Denmark, 2800 Kongens Lyngby, Denmark}

\begin{abstract}
We establish limitations and bounds on the transmission of quantum states between gravitationally interacting mechanical oscillators under different models of gravity. This provides benchmarks that can enable tests for quantum features of gravity. Our proposal does not require the measurement of gravitationally induced entanglement and only requires final measurements of a single subsystem. We discuss bounds for classical models based on local operations and classical communication when considering coherent-state alphabets, and we discuss the transfer of quantum squeezing for falsifying the Schrödinger-Newton model.
\end{abstract}

\maketitle

\section{Introduction}
The quest to formulate a consistent ultraviolet-complete theory of quantum gravity has haunted physicists for the last century. So much so, that the multitude of theoretical obstacles faced in pursuing such a theory has led some to question the very assumption that the gravitational interaction needs to be quantized. Fueled by the incredible pace at which quantum control and ground-state cooling techniques are developing, several experimental proposals have been put forward in recent years to answer this simple question: \textit{is the gravitational interaction fundamentally quantum?} Of the many experimental proposals for the detection of quantum effects in gravity, lots of attention has focused on experiments that look for so-called gravity-induced entanglement (GIE), which is seen as proof (or, at least, evidence) of the quantum nature of gravity \cite{Bose2017, Marletto2017}. 

\begin{figure}[h!]
 \centering
 \includegraphics[width=1\linewidth]{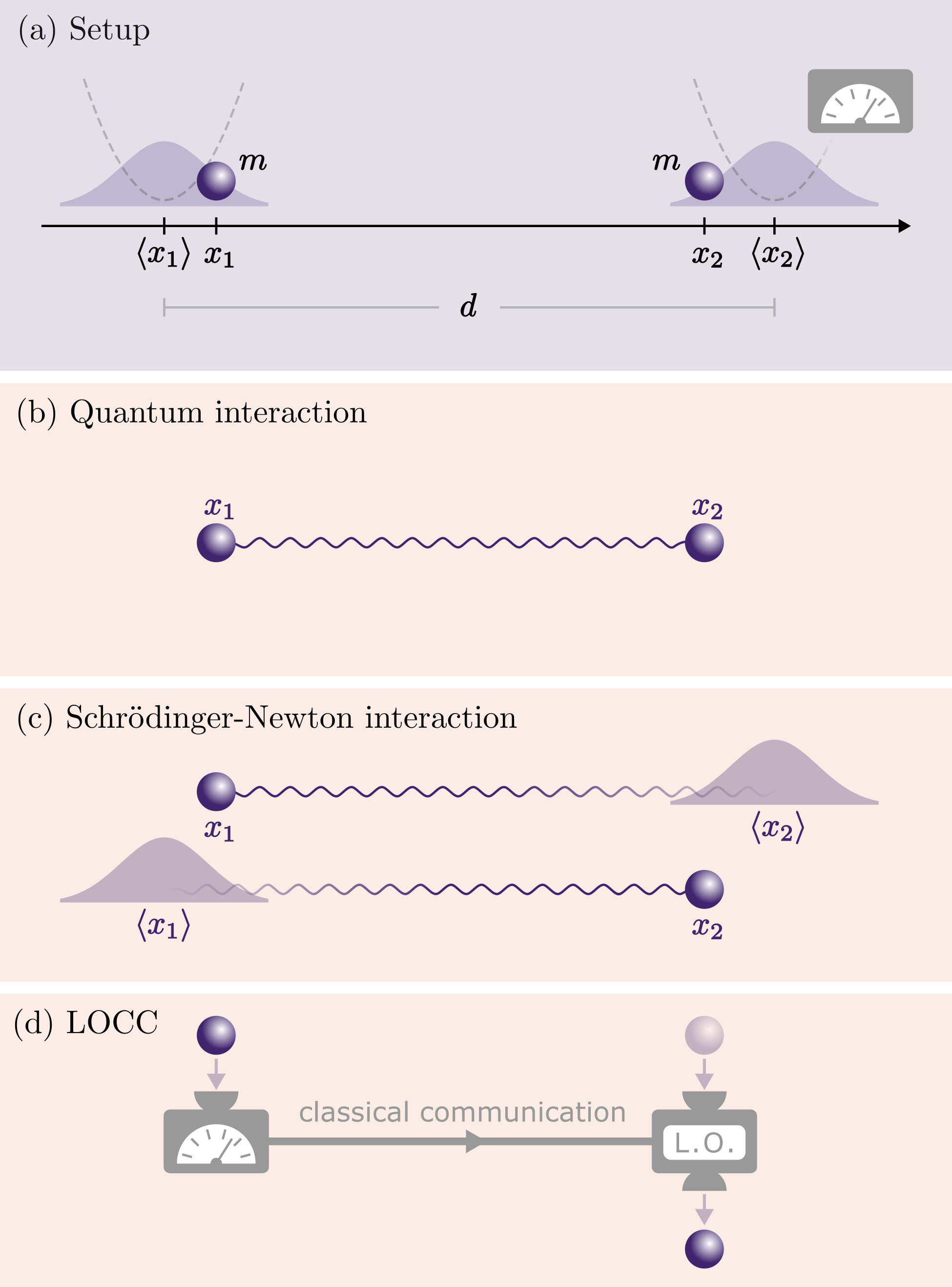}
 \caption{\textbf{(a)} 
 Two identical masses $m$ in local harmonic potentials interact through gravity. The separation $d$ between the equilibrium positions of each oscillator is much larger than the fluctuations. The systems interact for some time, at which point a measurement is performed on one of the oscillators. \textbf{(b)-(c)} Different models of the gravitational interaction. \textbf{(b)} The \textit{null hypothesis} described by the Newtonian Hamiltonian \eqref{eq:newtonian-interaction}. \textbf{(c)} In the Schrodinger-Newton model each system sees the gravitational effect of the other as a driving force dependent only on the average position. \textbf{(d)} A LOCC model that attempts to simulate the null hypothesis quantum dynamics. Under this model, gravity is an entity that performs local measurements on each system and communicates the outcomes to the other. }
\label{fig:concept}
\end{figure}

Pivotal to these proposals, referred to as Bose-Marletto-Vedral (BMV) experiments, is the idea that the gravitational interaction between quantum systems can be understood as a channel -- in the quantum information theoretic sense, a completely positive and trace preserving (CPTP) map. The channel implementing the gravitational interaction is deemed a classical channel if realized through local operations and classical communication (LOCC). On this basis, the gravitational interaction cannot be classical if observed to entangle product states, given that LOCC channels cannot map separable to entangled states \cite{Horodecki2009}. A number of other experimental proposals to test quantum aspects of gravity also exist \cite{Belenchia2018, Howl2021, Haine2021, Carlesso2019, Schmöle2016, Krisnanda2019, Carney2021, Qvarfort2020, Datta2021, Matsumura2022}. We will also mention that the publication of \cite{Bose2017, Marletto2017} has ignited a lively debate concerning the correct interpretation of the potential experimental results of BMV-like experiments spanning both the quantum gravity and quantum information communities \cite{Carney2022, Fragkos2022, Danielson2022, Christodolu2023b, Christodoulou2023a, Martinez2023, Döner2022, Bose2022}. The realization of BMV experiments also faces numerous challenges from an experimental point of view. Firstly, every interaction, except the gravitational one, must be shielded for the BMV proposals to make sense. Most importantly, the protocols assume the ability to generate macroscopic spatial superpositions of large masses, such as cat states, and to keep such superpositions coherent long enough for the induced entanglement to be appreciable. This is a Herculean feat since such experiments are inherently marred by decoherence \cite{Bose2017, Bose2024, Chevalier2020, Guff2022, Vandekamp2020, Fragolino2024, Toroš2021, Pedernales2020, Weiss2021, Rijavec2021, Gunnink2023}. We refer to the reviews \cite{Bose2024, Marletto2024} for a more complete rundown of the experimental challenges. 

However, a recent proposal \cite{Lami2024} promises to alleviate the challenges related to generating highly localized macroscopic spatial superpositions of large masses with a protocol that circumvents the need for such states in the first place. Their simple idea relies on the fact that the set of non-LOCC operations is larger than the set of entangling operations. In other words, quantum interactions can do more than entangle product states. To illustrate this fact, the authors consider the swap unitary, namely, given two states $\ket{\psi}\in\mathcal{H}_1$ and $\ket{\varphi}\in\mathcal{H}_2$ the unitary $S_{12}(\ket{\psi}_1\otimes\ket{\varphi}_2)=\ket{\varphi}_1\otimes\ket{\psi}_2$. Given these states, $S_{12}$ does not generate entanglement, but at the same time, it is non-LOCC. Then, if an unknown interaction is shown to swap states as illustrated above, such interaction is necessarily not LOCC. Building on this insight, the authors \cite{Lami2024} propose quantitative 
 constraints on all the possible LOCC dynamics. They call these \textit{LOCC inequalities}. These inequalities quantify how well LOCC models can simulate known quantum unitaries -- such as the unitary generated by the quantum gravitational dynamics. Crucially, they show that violating these inequalities in the gravitational context can be achieved with states that are relatively easy to prepare in experimental settings, that is, with coherent states. 

 In this work, we propose a simplified version of the protocol presented in \cite{Lami2024}. Considering two gravitationally interacting harmonic oscillators, we find tight LOCC fidelity bounds for the scenario where the input states of one oscillator are randomly picked from a prior probability distribution, the other oscillator is prepared in a reference state, and output measurements are only possible on one of the interacting sub-systems. This differs from the entanglement-witness \cite{Bose2017, Marletto2017} and the Lami et al. \cite{Lami2024} protocols, which require final measurements on both interacting sub-systems. The bounds are quantitative measures of the quantum communication capabilities of the gravitational interaction. A bound violation would indicate the gravitational interaction's ability to operate as a quantum communication channel. However, given the particular set of states assumed for the computation of the bound, namely a coherent-state alphabet, we will see that a LOCC violation does not rule out non-linear semi-classical models such as the Schrödinger-Newton (SN) model \cite{Gollapudi2024}. Nevertheless, we additionally show how such non-linear models can be ruled out for different sets of input states. Indeed, as we will articulate below, the Schrödinger-Newton model fails to transfer squeezing from one oscillator to the other.  

The rest of the paper is organized in the following way. In \secref{sec:system}, we introduce some notation and the physical system of interest, i.e., two gravitationally coupled quantum oscillators -- see \figref{fig:concept}. In \secref{sec:gravitational-state-swap}, we illustrate how quantum states are swapped in the quantum gravitational model of the interaction and contrast this to what happens in the Schrödinger-Newton model. In \secref{sec:classical-simulation-bounds}, we review the LOCC inequalities mentioned above in more detail and determine the classical simulation bounds for two interacting harmonic oscillators when measurements are only carried out on a single subsystem. In \secref{sec:noise}, we discuss how unavoidable imperfections affect the protocols and discuss experimental implementations and the relationship between LOCC bounds and the Schrödinger-Newton model. Finally, we conclude in \secref{sec:conclusions}. 

\section{System -- two coupled oscillators}
\label{sec:system}
We consider a system composed of two identical oscillators of mass $m$ and frequency $\omega$ separated by a mean distance $d$ and coupled through gravity, as illustrated in \figref{fig:concept}(a). In the weak-field limit $m/d\ll m_\text{P}/\ell_\text{P}$, where $m_\text{P}$ and $\ell_\text{P}$ are the Planck mass and length, the systems interact through the Hamiltonian \cite{Carney2018}
\begin{align}
\label{eq:newtonian-interaction}
    \hat{H}_{\g} = - \frac{G  m^2 }{|d+\hat{x}_1-\hat{x}_2|}\, ,
\end{align}
where $G$ is Newton's constant. We will henceforth refer to the evolution generated by the Hamiltonian \eqref{eq:newtonian-interaction} as the \textit{null-hypothesis} quantum gravitational evolution.
Assuming that the quantum fluctuations around the equilibrium point of each oscillator are much smaller than the separation between the two oscillators, i.e., $d\gg | x_1 - x_2|$, we expand the interaction term, leading to the Hamiltonian \cite{Carney2018}
\begin{align}
\label{eq:coupled-oscill}
    \hat{H}_\g = \hbar \,\gamma_\g\left( \hat{a}_1 \hat{a}_2^\dagger + \hat{a}_1^\dagger \hat{a}_2 + \hat{a}_1 \hat{a}_2 + \hat{a}_1^\dagger \hat{a}_2^\dagger \right) \, ,
\end{align}
with $\gamma_\g = {G m}/{\omega d^3}$ and $\hbar$ the reduced Planck constant. The mode operators are related to the position and momentum by
\begin{align}
    \hat{x}_k=\sqrt{\frac{\hbar}{2m\omega}}\left(\hat{a}_k^\dagger + \hat{a}_k\right), \quad \hat{p}_k= i\sqrt{\frac{\hbar m\omega}{2}}\left(\hat{a}_k^\dagger - \hat{a}_k\right) \, ,
\end{align}
where $k\in\{1,2\}$. We can further approximate the interaction Hamiltonian using the rotating wave approximation (RWA), which is valid if the frequency of each oscillator $\omega$ is larger than the interaction strength $\gamma_\g$ -- a reasonable approximation considering the typically minuscule interaction strength. Therefore, the full Hamiltonian reads 
\begin{align}
\label{eq:RWA-hamiltonian}
    \hat{H} = \hbar\, \omega \,\hat{a}_1^\dagger\hat{a}_1 + \hbar\, \omega \, \hat{a}_2^\dagger\hat{a}_2 + \hbar \,\gamma_\g\left( \hat{a}_1 \hat{a}_2^\dagger + \hat{a}_1^\dagger \hat{a}_2  \right) \, .
\end{align}
With these approximations, the gravitational interaction is equivalent to a beam splitter transformation, and the system dynamics can be solved easily in the Heisenberg picture, where we find that
\begin{align}
\begin{split}
   & \hat{a}_1^\dagger(t) = e^{i\omega t} \left(\hat{a}_1^\dagger(0) \cos(\gamma_\g t)+ i \hat{a}_2^\dagger(0) \sin(\gamma_\g t)\right), \\ & \hat{a}^\dagger_2(t) = e^{i\omega t} \left(\hat{a}_2^\dagger(0) \cos(\gamma_\g t)+ i \hat{a}_1^\dagger(0) \sin(\gamma_\g t)\right) \, .
\end{split}
\end{align}

In the following, we will compare the fully quantum beam-splitter-like interaction above to classical models of gravity in terms of their ability to transfer quantum characteristics from one oscillator to the other.

\section{Gravitational state swapping}
\label{sec:gravitational-state-swap}
We first compare the quantum gravitational interaction to the Schrödinger-Newton model \cite{Diosi1984,Penrose1996}. The central difference is illustrated in \figref{fig:concept}(b) and (c). In contrast to the quantum interaction, in the Schrödinger-Newton model, each oscillator interacts only with the average position of the other. This limits the possibilities for transferring quantum states between them.

Consider first the quantum case and let us assume the system is initially prepared in the state
\begin{align}
    \ket{\Psi(0)} = \ket{\psi} \otimes \ket{\phi}
\end{align}
where $\ket{\psi}$ and $\ket{\phi}$ are general states of the first and second oscillators. Since the gravitational interaction acts as a beam splitter, it is easy to see that for times $t_\text{s}=(2k+1)\, \pi / (2\gamma_\g)$ with $k\in \mathbb{N}$, the system is in the state 
\begin{align}
    |\Psi(t_\text{s})\rangle=|\tilde{\phi}\rangle\otimes|\tilde{\psi}\rangle,
\end{align}
where, given the phase-shift unitary $\hat{U}_i(\theta) = \exp(-i\theta\hat{n}_i)$, with $\hat{n}_i=\hat{a}^\dagger_i \hat{a}_i$, $\theta\in\mathbb{R}$, and $i\in\{1,2\}$, we have $|\tilde{\psi}\rangle= \hat{U}_1({\omega t_\text{s}})|\psi\rangle$ and $|\tilde{\phi}\rangle=\hat{U}_2(\omega t_\text{s})|\phi\rangle$. The low-energy quantum gravitational interaction, in the limit described by the Hamiltonian \eqref{eq:RWA-hamiltonian}, thus enables complete quantum state transfer between the two oscillators (up to local phases) \cite{Lami2024}. For instance, squeezing can be transferred from one mode to the other, which is evident from how the rotated quadratures $\hat{x}_i(\theta) = \hat{U}^\dagger _i(\theta)\hat{x}_i \hat{U}_i(\theta)$ transform after the system has interacted for $t=t_\text{s}$
\begin{align}
\begin{split}
    &\hat{x}_1(\theta)\rightarrow \hat{U}^\dagger_2(\theta+\omega t_\text{s}) \, \hat{x}_2 \,\hat{U}_2(\theta +\omega t_\text{s}),\\
    &\hat{x}_2(\theta)\rightarrow \hat{U}^\dagger_1(\theta+{\omega t_\text{s}}) \, \hat{x}_1 \hat{U}_1(\theta+{\omega t_\text{s}}).
\end{split}
\end{align}
Hence, if the quadrature $x_1(\theta)$ is at $t=0$ squeezed along the direction specified by the squeezing angle $\varphi\in\mathbb{R}$, at $t=t_\text{s}$ the quadrature $x_2(\theta)$ will be squeezed by the same amount along the direction specified by the angle $\tilde{\varphi}=\varphi+\omega t_\text{s}$, namely
\begin{align}
    \Delta x_1(\varphi)|_{t=0} = \Delta x_2(\tilde{\varphi})|_{t=t_\text{s}}=\sqrt{\frac{\hbar}{2 m \omega}}e^{-2s},
\end{align}
where $s \in \mathbb{R}$ is the squeezing factor, and $e^{-2s}$ represents the squeezing magnitude in the $\hat{x}_1(\theta)$ quadrature at $t = 0$. Already at intermediate times $0<t<t_s$, some amount of squeezing will be present in the quadrature $\hat{x}_2$. We can contrast this to what happens if the gravitational interaction is modeled according to the Schrödinger-Newton equation \cite{Bahrami2014, Yang2013}. In the limit where $d \gg | x_1 - x_2|$, the Shrödinger-Newton interaction Hamiltonian reads \cite{Yang2013}
\begin{align}
\label{eq:SN-interaction}
\begin{split}
    \hat{H}_\g^\text{SN} =- C_1 &(\hat{x}_1 - \hat{x}_2) \\& - \frac{C_2}{2} \left[ \left(\hat{x}_1-\langle\hat{x}_2\rangle\right)^2+\left(\hat{x}_2-\langle\hat{x}_1\rangle\right)^2\right],
\end{split}
\end{align}
where $C_1 = G m^2 / d^2$ and $C_2 = G m^2 / d^3$. Let us suppose the initial state is the product state
\begin{align}
\label{eq:initial-state-SN}
    |\Phi(0)\rangle = |\xi \rangle \otimes |0\rangle
\end{align}
where $\ket{0}$ is the ground state, and $\ket{\xi}$ is the single mode squeezed vacuum state
\begin{align}
    |\xi\rangle =\frac{1}{\sqrt{\cosh(s)}} \sum_{n=0}^{\infty}(e^{-i\varphi}\tanh(s))^n\frac{\sqrt{(2n)!}}{2^n n!}|2n\rangle,
\end{align}
with squeezing parameter $s$, and squeezing angle $\varphi$. We may solve for the expectation values $\langle \hat{x}_1\rangle$ and $\langle \hat{x}_2 \rangle$ -- see \appref{appendix:SN-model} for a derivation -- and recast the Schrödinger-Newton Hamiltonian in the form
\begin{align}
    \hat{H}^\text{SN} =\sum_{k=1}^2\left( \frac{\hat{p}_k^2}{2m}+ \frac{1}{2} m \tilde{\omega}_\g^2\hat{x}^2_k-\hat{x}_k J(t)\right),
\end{align}
where
\begin{align}
  J(t) =C_1 - \frac{ C_1}{m\omega_\g^2}\left(1-\cos(\omega_\g t)\right),
\end{align}
$\omega_\g = \sqrt{\omega^2-2C_2/m}$, and $\tilde{\omega}_\g = \sqrt{\omega^2-C_2/m}$. Thus, in the semi-classical Schrödinger-Newton model, the two interacting oscillators see each other's gravitational effect as a classical driving force. Solving the dynamics in the Heisenberg picture, we find that evolving from the initial state \eqref{eq:initial-state-SN} results in
\begin{align}
    \Delta \hat{x}_2(t) = \Delta \hat{x}_2(0)=\sqrt{\frac{\hbar}{2 m \omega}}, \quad \forall\, t\in[0,\infty) ,
\end{align}
and so there will be no transfer of squeezing in the 
Schrödinger-Newton model. An experiment in which the transfer of squeezing from one oscillator to the other is observed could thus falsify the Schrödinger-Newton model.

\section{Classical simulation bounds}
\label{sec:classical-simulation-bounds}
The results above show that the Schrödinger-Newton model may be falsified by observing the transfer of quantum features, such as squeezing, through the gravitational interaction. However, Schrödinger-Newton is not the only possibility for a fundamentally classical description of gravity. An alternative approach to testing for quantum features of the gravitational interaction is to compare the quantum predictions against general models that assume gravity acts as a LOCC channel, as illustrated in \figref{fig:concept} (d). In other words, the question we want to answer is, to what extent can a LOCC channel simulate the null-hypothesis quantum model? 

\subsection{LOCC simulation fidelity bound}
A way of quantifying how well a LOCC channel can simulate a unitary transformation $U$ was introduced by Lami et al. \cite{Lami2024}, in the form of what they call an \textit{LOCC inequality}. Let us briefly review their findings. 

Let $S=A_1 A_2$ be a bi-partite quantum system, and let $(U_t)_{t\in\mathbb{R}}$ be a strongly continuous one-parameter unitary group such that $U_t:\mathcal{H}_S\rightarrow\mathcal{H}_{S}$ are unitary transformations on the Hilbert space $\mathcal{H}_S$ associated with the quantum system $S$. We adopt the standard nomenclature where $A_1$ is controlled by Alice and $A_2$ is controlled by Bob. Later, we will take $U_t$ to be the unitary that implements the quantum gravitational evolution. Given an ensemble of random pure states, defined by the states $\rho(\alpha)= |\psi_\alpha\lr \psi_\alpha|$ and the probability distribution $p(\alpha)$, from which we pick the states of the system $S$, the LOCC simulation fidelity bound is the maximal average fidelity between the target state $\rho^{}_t(\alpha)=U_t(\rho(\alpha))=U_t\, \rho(\alpha)\, U^\dagger_t$ and its simulation $\mathcal{E}(\rho(\alpha))$, where $\mathcal{E}$ is a LOCC channel on $S$, namely
\begin{align}
\label{eq:LOCC-bound}
    F_{c\ell}(t) =\sup_{\mathcal{E}\in\text{LOCC}}\sum_\alpha p(\alpha)\text{Tr}\left[\mathcal{E}(\rho(\alpha))\,\rho^{}_t(\alpha)\right].
\end{align}
To interpret the significance of the defined quantity, we consider the following scenario: Alice and Bob are given an unknown one-parameter channel that acts on $S$, and they suspect that the channel is implemented by $U_t$ for certain values $t\in\mathbb{R}$ -- we refer to this as the null hypothesis. Their goal is to certify the null hypothesis and make sure that the channel is not secretly a classical (LOCC) channel that simulates the unitary $U_t$. Therefore, the quantity \eqref{eq:LOCC-bound} sets a benchmark, which, if broken, ensures the non-LOCCness of the channel. In practice, an experimental protocol would follow the steps below.

\begin{enumerate}[(i)]
    \item The bi-partite system is initialized in a random pure state picked from the ensemble $\{\rho(\alpha),p(\alpha)\}$.
    \item The channel acts on the input states for a choice of the parameter $t\in\mathbb{R}$.
    \item The output state $\rho^{}_t(\alpha)$ is computed and the POVM measurement corresponding to $\{\rho^{}_t(\alpha),\id-\rho^{}_t(\alpha)\}$ is carried out on each system.
\end{enumerate}

By running steps (i) to (iii) a statistically significant number of times, a verifier can compare the measured outcomes with the LOCC bound \eqref{eq:LOCC-bound}. Supposing the channel is implemented by the null hypothesis, the verifier will always report a fidelity greater than the LOCC simulation bound. In particular, assuming ideal measurements and closed dynamics, the fidelity would be exactly $F_\text{measured}=1$. 

\begin{figure}[t]
    \centering
    \includegraphics[width=1\linewidth]{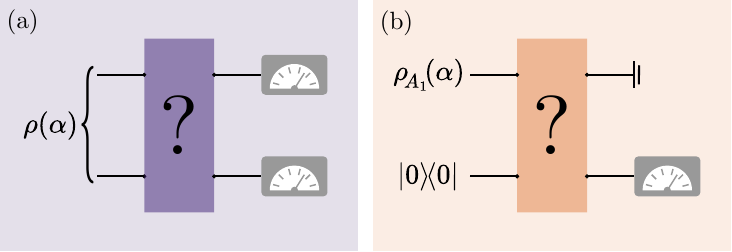}
    \caption{\textbf{(a)} Schematic representation of the protocol proposed in \cite{Lami2024}. Two quantum systems $A_1$ and $A_2$ are initialized in a random pure state picked from the ensemble $\{\rho(\alpha),p(\alpha)\}$. The state of the combined system is fed to an unknown channel of which we want to determine the LOCCness. In the end, the binary measurement $\{\rho^{}_t(\alpha),\id-\rho^{}_t(\alpha)\}$ is performed on both systems. \textbf{(b)} Schematic representation of the protocol discussed in this paper. The first oscillator is prepared in a state picked from the ensemble $\{\rho^{}_{A_1}(\alpha),p(\alpha)\}$, while the second oscillator is in the vacuum. The states are fed to the unknown channel as above. At the end, the projective measurement $\{{\rho^{}_{t,A_2}}(\alpha),\id-{\rho^{}_{t,A_2}}(\alpha)\}$ is carried out on the second sub-system.}
    \label{fig:two_vs_one_measurment_schemes}
\end{figure}

We will be interested in a slightly different bound -- see \figref{fig:two_vs_one_measurment_schemes} for a pictorial representation. Suppose we are only allowed to carry out measurements on one of the output sub-systems, say Bob's, we can still formulate a bound analogous to \eqref{eq:LOCC-bound} when such restriction is in place. Suppose that Alice's sub-system is initialized in a random pure state picked from the ensemble $\{\rho^{}_{A_1}(\alpha),p(\alpha)\}$, while Bob initializes his state to the vacuum $|0\lr0|$. As above, we denote with $\rho(\alpha)$ the full state of the system $S=A_1 A_2$, and with $\rho^{}_t(\alpha)=U_t(\rho(\alpha))$ the output state. We can run the following protocol to verify the quantumness of the interaction between $A_1$ and $A_2$.

\begin{enumerate}[(1)]
    \item Alice's state is initialized in a random pure state picked from the ensemble $\{\rho^{}_{A_1}(\alpha),p(\alpha)\}$ while Bob's subsystem is initialized to the vacuum.
    \item The channel acts on the input states for a choice of the parameter $t\in\mathbb{R}$.
    \item Bob's output state $\rho^{}_{t,A_2}(\alpha)=\text{Tr}_{A_1}[\rho^{}_t(\alpha)]$, where $\text{Tr}_{A_1}[\,\cdot\,]$ stands for the partial trace over Alice's system, is computed and the POVM measurement corresponding to $\{{\rho^{}_{t,A_2}}(\alpha),\id-{\rho^{}_{t,A_2}}(\alpha)\}$ is carried out on Bob's sub-system.
\end{enumerate}

Similar to the above scenario, after running steps (1) to (3) a statistically significant number of times, we can compare the measured outcomes to the modified LOCC bound 
\begin{align}
\label{eq:general-teleportation-bound}    F_{c\ell}(t)=\sup_{\mathcal{E}\in\text{LOCC}}\sum_\alpha p(\alpha)\text{Tr}\left[\mathcal{E}(\rho^{}_{A_1}(\alpha)) \rho^{}_{t,A_2}(\alpha)\right] ,
\end{align} 
if the bound is violated, the interaction could not have happened over a LOCC channel. We will now apply these considerations in the context of gravitational interactions.

\subsection{Quantum benchmark for gravitational transmission of coherent states}
\begin{figure}[t!]
    \centering
    \includegraphics[width=1\linewidth]{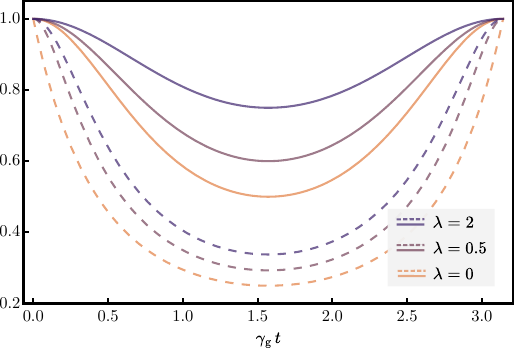}
    \caption{The classical fidelity bound \eqref{eq:ourlocc-bound} (solid) and the LOCC simulation bound (dotted) from \cite{Lami2024} vs.\ time for different $\lambda$. As expected, the bound \eqref{eq:ourlocc-bound} is looser than the bound of \cite{Lami2024} because information is lost when tracing out Alice's system. Simulating the reduced rather than the full dynamics is easier for the classical model.}
    \label{fig:bound-plots}
\end{figure}

Let us return to the two gravitationally interacting oscillators.  Following the notation from the previous section, we have $S=A_1 A_2$. We will now compute the simulation bound \eqref{eq:general-teleportation-bound} in the case where the initial state of Alice's system is in a random coherent state. While it is hard to characterize general LOCC operations mathematically, we are only interested in LOCC simulations of processes that ultimately map $A_1\rightarrow A_2$. Therefore, we can characterize such $\text{LOCC}(A_1\rightarrow A_2)$ channels with measure-and-prepare channels of the kind
\begin{align}
    \mathcal{E}(\rho)=\sum_\mu \tr\left[\rho\,\Pi_\mu\right]\rho^{}_\mu,
\end{align} 
where $\{\Pi_\mu\}$ is a positive operator-valued measurement (POVM) and $\rho$ is a state. This characterization will allow us to determine the exact bound \eqref{eq:general-teleportation-bound}. 
We consider the initial state of the system $S$, to be in the pure state
\begin{align}
\label{eq:initial-state}
    |\Psi(0)\rangle = |\alpha\rangle \otimes | 0 \rangle.
\end{align} 
The state of the first oscillator $A_1$ is picked from the pure state ensemble $\{|\alpha\lr\alpha|,p(\alpha)\}$, where $\alpha\in\mathbb{C}$, 
\begin{align}
\label{eq:gauss-distribution}
    p(\alpha)= \frac{\lambda}{\pi}\,e^{-\lambda|\alpha|^2}\, ,
\end{align}
and $\lambda>0$ is the inverse of the distribution width, while the second oscillator $A_2$ is prepared in the vacuum. Under the null-hypothesis quantum gravitational dynamic 
\begin{align}
\label{eq:gravity-unitary}
    U_t=e^{-i\gamma_\g t (a_1 a_2^\dagger + a_1^\dagger a_2)},
\end{align}
the initial state \eqref{eq:initial-state} evolves to the product state
\begin{align}
    |\Psi(t)\rangle = |\alpha \cos (\gamma_\g t)\rangle \otimes |-i\alpha \sin(\gamma_\g t) \rangle \, ,
\end{align}
where we ignore the phase terms originating from the free evolution. As discussed above, for $t_\text{s}= \pi / (2\gamma_\g)$ Alice's and Bob's states are fully swapped. Hence, Alice can use the gravitational interaction to transmit her state to Bob. Evidently, if the underlying gravitational interaction acted as a LOCC channel such transmission would not be possible. The classical fidelity bound for a quantum teleportation channel is the well-known result \cite{Hammerer2005}
\begin{align}
    F_{c\ell}\leq \frac{1 + \lambda}{2+ \lambda}\, .
\end{align}
However, we want to know the classical bound under the time evolution generated by the unitary \eqref{eq:gravity-unitary} in the interval $0\leq t \leq t_\text{s}$. Given that Bob's state at time $t$ is $|\alpha\sin(\gamma_\g t)\lr\alpha\sin(\gamma_\g t)|$, the bound we are looking for is given by 
\begin{align}
    F_{c\ell}(t)= \sup_{\mathcal{E}}\sum_\alpha p(\alpha)\langle\alpha\sin(\gamma_\g t)|\mathcal{E}(|\alpha\rangle\!\langle\alpha|)|\alpha\sin(\gamma_\g t)\rangle 
\end{align}
In other words, the classical channel has to simulate the physical process $|\alpha\rangle \mapsto|\alpha \sin(\gamma_\g t)\rangle$. The classical fidelity threshold for the physical process $|\alpha\rangle\mapsto |g \alpha \rangle$ was calculated for $g>1$ in \cite{Chiribella2013}, and is given by
\begin{align}
\label{coherent-ampl}
    F_{c\ell}=\frac{1+\lambda}{1+\lambda+g^2}.
\end{align}
In our case, $g=\sin(\gamma_\g t)$, but the result \eqref{coherent-ampl} can be extended to $0\leq g\leq1$ -- see \appref{appendix:bound-proof}. Hence, the bound we are looking for is 
\begin{align}
\label{eq:ourlocc-bound}
       F_{c\ell}(t)=\frac{1+\lambda}{1+\lambda+\sin^2(\gamma_\g t)}.
\end{align}
A plot of the above bound for various choices of the inverse distribution width $\lambda$  is presented in \figref{fig:bound-plots}.

\section{Imperfections and implementation}
\label{sec:noise}
\subsection{Noisy dynamics}
Any experimental implementation will be subject to environmental noise, leading to decoherence. In particular, thermal noise and mechanical damping of the oscillator's motion can significantly affect the protocol. Assuming this noise and damping to be Gaussian, since the interaction Hamiltonian \eqref{eq:RWA-hamiltonian} is quadratic, we can describe the evolution through the first and second moments using the symplectic formalism \cite{Serafini2017}. For convenience, we set $\hbar=1$ in this subsection. Given the vector of quadrature operators $\boldsymbol{r}=(x_1,p_1,x_2,p_2)^{\text{T}}$, we can define the covariance matrix and displacement vector for the two oscillators by
\begin{align}
    \sigma_{kl}=\frac{1}{2}\langle\{r_k,r_l\}\rangle-\langle r_k\lr r_k\rangle , \quad \bar{r}_k=\langle r_k \rangle,
\end{align}
where $\{\cdot,\cdot\}$ is the anticommutator, and $\langle A \rangle = \text{Tr}[\rho A]$. 

Under Markovian evolution, the covariance matrix evolves according to the evolution equations \cite{Serafini2017}
\begin{align}
\label{eq:open-dynamics}
\begin{split}
    \frac{\dd}{\dd t}\boldsymbol{\sigma}&=A \boldsymbol{\sigma} + \boldsymbol{\sigma}A^\text{T}+D, \\
    \frac{\dd}{\dd t}\bar{\boldsymbol{r}}&= A \bar{\boldsymbol{r}} + \boldsymbol{d}
\end{split}
\end{align}
where $A=\boldsymbol{\Omega}H-\frac{\gamma}{2}\,\id$ is the drift matrix, $\gamma$ the oscillator decay rate, $\boldsymbol{\Omega}$ the symplectic matrix, $D=(2 \bar{N}+1)\gamma\,\id$, and  $\bar{N}$ the average number of thermal phonons in the environment \cite{Serafini2017}. The Hamiltonian matrix is explicitly given by
\begin{align}
     H&=\gamma_\g\begin{pmatrix}
     \,0& 0 & 1 & 0\,\\
     \,0 & 0 & 0 & 1\,\\
     \,1 & 0 & 0 & 0\,\\
     \,0 & 1 & 0 & 0\,
\end{pmatrix}.
\end{align}
 Let us assume that the preparation of the initial state \eqref{eq:initial-state} is not perfect and subject to thermal noise, so that the initial state corresponds to displacing the thermal state
\begin{align}
\label{eq:initial-cov}
    \boldsymbol{\sigma} (0)&=\begin{pmatrix}
    \bar{n}+\frac{1}{2} & 0 & 0 & 0\\
    0 & \bar{n}+\frac{1}{2} & 0 & 0\\
    0 & 0 & \bar{n}+\frac{1}{2} & 0\\
    0 & 0 & 0 & \bar{n}+\frac{1}{2} 
\end{pmatrix},
\end{align}
as opposed to the vacuum state. Here, $\bar{n}$ is the average initial thermal phonon number and the displacement vector is given by
\begin{align}
\label{eq:initial-disp}
\bar{\boldsymbol{r}}(0)&=\sqrt{2}\begin{pmatrix}
    \text{Re}(\alpha)\\
    \text{Im}(\alpha)\\
    0 \\
    0  
\end{pmatrix}.
\end{align}

\begin{figure}[t!]
    \centering
    \includegraphics[width=1\linewidth]{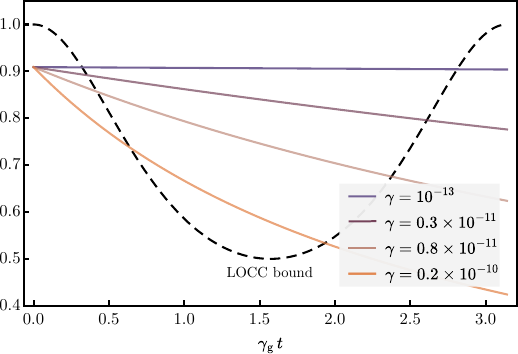}
    \caption{Plots of the overlap \eqref{eq:fidelity-open} between the closed and open dynamics output state, for different values of the decay rate $\gamma$. The LOCC bound corresponds to the choice of $\lambda=10^{-3}$ for the prior distribution. We fixed $\bar{N}=10^{10}$ and $\bar{n}=10^{-1}$.}
    \label{fig:open-dynamics-1}
\end{figure}

Then, under the Markovian evolution \eqref{eq:open-dynamics} at time $t$ the covariance matrix and displacement vector of Bob's subsystem will evolve to
\begin{align}
\label{eq:noisy-evolution-states}
\begin{split}
        &\boldsymbol{\sigma}_{A_2}(t)=e^{-\gamma t} \boldsymbol{\sigma}_{A_2}(0)+ 
    \left(1 + 2 \bar{N} (1-e^{-\gamma t})\right)\id
    \\
    &\bar{\boldsymbol{r}}_{A_2}(t)=\sqrt{2}\,  e^{-\frac{\gamma t}{2}} \sin (\gamma_\g  t) \,\Bigl(\text{Im}(\alpha),-\text{Re}(\alpha) \Bigr)^\text{T}
\end{split}
\end{align}
The overlap between two Gaussian states $\rho^{}_1$ and $\rho^{}_2$ (of which at least one is pure) in terms of their displacement vector and covariance matrix is given by
\begin{align}
\label{eq:gaussian-states-overlap}
    F(\rho^{}_1,\rho^{}_2)=\frac{1}{\det(\boldsymbol{\sigma_1}+\boldsymbol{\sigma_2})^\frac{1}{4}} e^{-\frac{1}{4}(\bar{\boldsymbol{r}}_1+\bar{\boldsymbol{r}}_2)^\text{T} (\boldsymbol{\sigma_1}+\boldsymbol{\sigma_2})^{-1} (\bar{\boldsymbol{r}}_1+\bar{\boldsymbol{r}}_2)  }.
\end{align}
From \eqref{eq:noisy-evolution-states} and \eqref{eq:gaussian-states-overlap}, the fidelity of the noisy to the ideal output state can be determined readily -- the result will depend on $\alpha$, so one should take care of averaging over the prior distribution \eqref{eq:gauss-distribution}. However, if the decay rate $\gamma$ is known, the overlap can be improved by compensating for the displacement due to the lossy dynamics. This would amount to applying the displacement
\begin{align}
\label{eq:conpensation-displacement}
    D_{\tilde{\alpha}}=e^{\left(\tilde{\alpha}\, \hat{a}_2^\dagger-\tilde{\alpha}^*\hat{a}_2\right)}, \quad \tilde{\alpha}=-i\,\alpha\sin (\gamma_\g  t)\left(1-e^{-\frac{\gamma t}{2}}\right),
\end{align}
before performing the final projective measurement. Thus, the displacement vector \eqref{eq:noisy-evolution-states} is mapped to
\begin{align}
    D_{\tilde{\alpha}}^\dagger\, \bar{\boldsymbol{r}}_{A_2} D_{\tilde{\alpha}}=\sqrt{2}\, \sin (\gamma_\g  t) \,\Bigl(\text{Im}(\alpha),-\text{Re}(\alpha) \Bigr)^\text{T}.
\end{align}
while the covariance matrix is unchanged. Therefore, the overlap between open and closed dynamics evaluates to
\begin{align}
\label{eq:fidelity-open}
    F_\text{open}(t)=\frac{2 e^{\gamma  t}}{2 \bar{n}+(4 \bar{N}+3) e^{\gamma  t}-4 \bar{N}-1} .
\end{align}
In \figref{fig:open-dynamics-1}, the fidelity \eqref{eq:fidelity-open} is compared to the LOCC bound \eqref{eq:ourlocc-bound} computed in the previous section. As expected, even as we partially compensate for the lossy dynamics by displacing the second oscillator by \eqref{eq:conpensation-displacement}, we observe loss of coherence at a rate determined by the thermal occupation number $\bar{N}$ and the phonon dissipation rate $\gamma$.

\subsection{Some experimental considerations}
Let us discuss how the protocol above might be implemented in practice. As mechanical oscillators, we consider two suspended mirrors whose motion is controlled through optical cavities of high quality factor \cite{miao2020}. The gravitational interaction strength of the two mirrors sets the time scale of the experiment; assuming the mirrors are composed of a material with constant density $\varrho$, the interaction strength is \cite{miao2020}
\begin{align}
    \label{eq:interaction-strenght}
    \gamma_\g=\frac{\Lambda G \varrho}{\omega},
\end{align}
where $\Lambda$ is a geometric factor that accounts for the shape of the two mirrors and the center of mass separation $d$. As in \cite{miao2020}, we will assume that the mirrors have a cylindrical geometry, with radius $R$ and thickness $L$. When $R/L=3/2$, and $R=d$, the geometric factor evaluates numerically to $\Lambda\simeq2$ \cite{miao2020}. To estimate an upper bound on the interaction strength, we can assume the mirrors are made of the densest material known, osmium, which has a density of roughly $\varrho= 2.26\times 10^4\,\text{kg}/\text{m}^3$. 
Thus, for mechanical oscillators of frequencies $\omega =1.00\times10^{-2}\, \text{Hz}$ we obtain $\gamma_\g\simeq4.74\times10^{-4}\, \text{Hz}$. As we can see, these parameters are consistent with the rotating wave approximation that led us to the interaction Hamiltonian \eqref{eq:RWA-hamiltonian}, that is, $\omega\gg\gamma_\g$. Furthermore, since the protocol assumes that we are preparing the first oscillator in a coherent state whose amplitude is picked from the Gaussian distribution \eqref{eq:gauss-distribution}, we need to make sure that the displacements are not so large as to violate the assumption $|\hat{x}_1-\hat{x}_2|\ll d$. This condition restricts how small $\lambda$ can be. Since the second oscillator is initialized in the vacuum state, we want $\langle \hat{x}_1 \rangle \ll d$, that is
\begin{align}
    \sqrt{\frac{2 \hbar}{m \omega\lambda}}\ll d\, .
\end{align}
With the abovementioned parameters and assuming the separation to be in the order $d\simeq200 \,\mu\text{m}$ (which is large enough to neglect the Casimir-Polder interaction), we get $\lambda\gg10^{-9}$. From the bound \eqref{eq:ourlocc-bound}, we find that the time it takes for the fidelity to drop to a lower value, say $\mathcal{F}\in[\mathcal{F}^{(\lambda)}_\text{min},1)$ where $\mathcal{F}^{(\lambda)}_\text{min}$ is the lowest value assumed by the LOCC bound for a given value of $\lambda$, is given by
\begin{align}
    t=\frac{1}{\gamma_\g}\sin^{-1}\sqrt{\frac{(1+\lambda)(1-\mathcal{F})}{\mathcal{F}}}.
\end{align}
Therefore, with $\lambda=10^{-3}$, in a single experimental run of our protocol, a drop in fidelity bound to $\mathcal{F}=0.9$ is achieved in $t=7.23\times10^{2} \, \text{s}$. 

Let us now consider how the noisy dynamics discussed in the subsection above affect our experimental considerations. We modeled the noisy Markovian dynamics of the system accounting for the initial average thermal occupation number $\bar{n}$, the average occupation number $\bar{N}$ of a thermal bath that is symmetrically coupled to the oscillators, and the phonon dissipation rate $\gamma=\omega/Q$, where $Q$ is the quality factor of the mechanical oscillators. Starting from the parameter $\bar{n}$, the overlap between a displaced vacuum state and a displaced thermal state for the same displacement is $1/(1+\bar{n})$, so to reach a target fidelity $\mathcal{F}^{(\lambda)}_\text{min}<\mathcal{F}<1$ we find the following restriction 
\begin{align}
\label{eq:bound-on-average-honon-number}
    \bar{n}<\frac{1-\mathcal{F}}{\mathcal{F}}.
\end{align}
For example, if we set the target fidelity to $\mathcal{F}=0.9$, this would imply $\bar{n}<1.1\times10^{-1}$. In turn, using the Bose-Einstein distribution
\begin{align}
    \bar{n}=\frac{1}{e^{\hbar\omega/k_B T_\text{eff}}-1},
\end{align}
where $T_\text{eff}=T_\text{env}/Q$, the restriction on $\bar{n}$ translates to restrictions on $Q$ (therefore, on $\gamma$) and $\bar{N}$. Considering a mechanical oscillator of frequency $\omega=10^{-2}\,\text{Hz}$ in a cryogenic environment with temperature $T_\text{env}=10^{-3}\,\text{K}$, reaching the required target initial phonon occupation number of $\bar{n}\sim10^{-1}$ means cooling the oscillator to an effective temperature $T_\text{env}/Q<3.3\times10^{-14}\, \text{K}$. Let us remark that the incredibly high quality factor of $Q=10^{11}$ required to reach such low effective temperatures is beyond current technical capabilities for the low-frequency oscillators we are envisioning here. On a final note, we want to point out that thanks to the trick of counter displacing the oscillators in response to the noise -- see Eq. \eqref{eq:conpensation-displacement} from the previous section and the plot in \figref{fig:open-dynamics-1} -- we can, at the cost of extending the time duration of each experimental run, set even lower fidelity targets. 

\begin{figure}[t]
    \centering
    \includegraphics[width=1\linewidth]{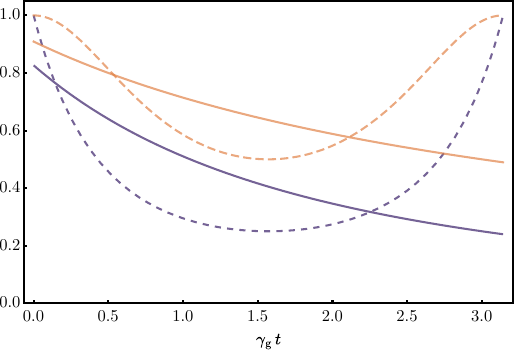}
    \caption{Plot comparing the fidelity of the open system dynamics -- as modeled by the Markovian evolution \eqref{eq:open-dynamics} -- to the closed quantum gravitational dynamics (solid curves) and the LOCC bound (dashed curves). The purple curves correspond to the proposal of Lami et al., whilst the orange curves correspond to our proposal. We chose $\lambda=10^{-3}$, $\bar{n}=10^{-1}$, $\bar{N}=10^{11}$, and $\gamma=1.5 \times10^{-12}$.}
    \label{fig:open_lami_v_us}
\end{figure}

\subsection{Comparison between measuring both or a single oscillator}
Our fidelity bound is looser compared to the bound established by Lami et al. \cite{Lami2024}  for two interacting oscillators (see \figref{fig:bound-plots}), which is expected. This is due to the information loss in our protocol when Alice’s system is traced out (see \figref{fig:two_vs_one_measurment_schemes}). As a result, a hypothetical LOCC model of gravity finds it easier to simulate the quantum channel. On the other hand, this could, in principle, mean that the setup of \cite{Lami2024} is more susceptible to noise and loss of quantum coherence. We can see this from the plot in \figref{fig:open_lami_v_us}. The fidelity between the ideal and noisy output state in the protocol of \cite{Lami2024} is well below the overlap curve for the single-measurement protocol -- where, as discussed in \secref{sec:classical-simulation-bounds}, we model the open dynamics using the evolution equations \eqref{eq:open-dynamics} which encode the coupling with a thermal environment and dampening of the oscillator's motion. That being said, it is important to keep in mind that we are only able to exclude perfect LOCC dynamics as long as the fidelity of the open--closed quantum gravitational dynamics is above the LOCC bound, namely, when we have $F_\text{open}(t)>F_{c\ell}(t)$, $\forall t \in (t_0, t_1)$. Our classical simulation fidelity bound \eqref{eq:ourlocc-bound} intersects the respective curve for the overlap between the open and closed dynamics at a later time $t_0$ compared to the corresponding intersection time $t_0'$ for the bound of \cite{Lami2024}. That being so, the overall lower values of the LOCC bounds in \cite{Lami2024} make up for the higher resilience to noise we might naively expect in our protocol. 

\subsection{Schrödinger-Newton violation of the LOCC bound}
\begin{figure}[t!]
    \centering
    \includegraphics[width=1\linewidth]{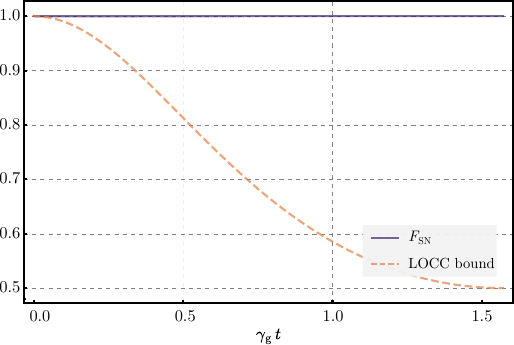}
    \caption{Plots comparing the LOCC bound \eqref{eq:ourlocc-bound} to the fidelity of the SN to the quantum gravitational evolution \eqref{eq:SN-quantum-fidelity}. The inverse width of the prior distribution is fixed to $\lambda=10^{-3}$. }
    \label{fig:SN-locc-violation}
\end{figure}

Before concluding, we would like to address the claim that the dynamics generated by the Schrödinger-Newton model fall under the classification of LOCC \cite{Lami2024}. It was recently pointed out that Schrödinger-Newton is not LOCC \cite{Gollapudi2024}. This is because any non-linear modification of the Schrödinger equation -- such as the SN equation -- leads to transformations that can map valid density matrix states to non-valid ones. In other words, the non-linear Schrödinger dynamic is not a CPTP map \cite{Gisin1989, Bahrami2014}.  In this sense, the transformations generated by Schrödinger-Newton dynamics would not even be classifiable as channels and, hence, cannot be LOCC. To corroborate this claim, we can verify whether the SN model violates the LOCC bound computed in this paper \eqref{eq:ourlocc-bound}. Accordingly, we compare the Schrödinger-Newton evolution of the random initial state $\ket{\Psi(0)}=\ket{\alpha}\otimes\ket{0}$, where $\alpha\in\mathbb{C}$ is picked from the i.i.d. distribution \eqref{eq:gauss-distribution}, to the quantum gravitational evolution \eqref{eq:gravity-unitary} by evaluating the average fidelity
\begin{align}
\label{eq:SN-quantum-fidelity}
    F_\text{SN}(t) = \int \dd^2\alpha \frac{\lambda}{\pi}  e^{-\lambda|\alpha|^2} \tr\left[\rho^{}_{t,A_2} \rho^{\text{SN}}_{t,A_2}(\alpha) \right],
\end{align}
where $\rho^{}_{t,A_2}(\alpha)$ and $\rho^{\text{SN}}_{t,A_2}(\alpha)$ are respectively the states of the second oscillator (Bob's state) after the quantum evolution \eqref{eq:gravity-unitary} and SN evolution \eqref{eq:full-SN-hamiltonian} in the limit $\omega\ll\gamma_g$. As we can see from \figref{fig:SN-locc-violation}, the LOCC bound is violated by the Schrödinger-Newton dynamics. Thus, if one is willing to see past other pathologies related to the Schrödinger-Newton model -- like superluminal signaling \cite{Gisin1989,Bahrami2014} -- and regards it as a potential candidate for a fundamental description of gravity, violation of the LOCC bound is not sufficient to rule it out and complementary tests, such as the one presented above, are necessary. If, on the other hand, one insists that a fundamentally classical theory of gravity must be LOCC, then Schrödinger-Newton cannot be a model of classical gravity.
\section{Conclusions}
\label{sec:conclusions}
In this article, we have discussed fundamental limits for classical models of gravity -- in the form of LOCC or Schrödinger-Newton models -- in transmitting quantum information. We have seen that if gravity is quantized in the Newtonian limit, two spatially separated parties controlling two harmonic oscillators can establish a quantum communication channel via the gravitational interaction, which allows the parties to exchange quantum states faithfully. In particular, preparing one oscillator in a certain state and subsequently measuring the other can result in behavior that cannot be reproduced by the classical models, thus allowing these models to be falsified. Our analysis demonstrates that observing the transfer of squeezing would falsify the Schrödinger-Newton model. Additionally, we derived a LOCC fidelity bound for a set of coherent states and compared it to scenarios where both oscillators are measured, as previously proposed. We have provided a thorough analysis of open-system dynamics, showing that although the parameters for potential implementations are demanding, the quantum behavior can persist in the presence of thermal noise and damping, providing hope for realizing experiments in the future.

\begin{acknowledgments}
We thank Alexssandre de Oliveira Junior, Konstantin Beyer, and Igor Pikovski for their valuable insights and comments. We gratefully acknowledge support from the Danish National Research Foundation, Center for Macroscopic Quantum States (bigQ, DNRF142), and Novo Nordisk Foundation (CBQS NNF 24SA0088433).
\end{acknowledgments}

\bibliography{gravityQC}

\appendix
\onecolumngrid
\section{Schrödinger-Newton interaction}
\label{appendix:SN-model}
In this appendix section, we derive the Schrödinger-Newton interaction potential \eqref{eq:SN-interaction} for two mesoscopic quantum systems. Starting from the interaction \cite{Bahrami2014}
\begin{align}
\label{eq:full-SN-potential}
    V_\text{SN}=G\,m^2\sum_{i,j=0}^2\int \dd x_1' \dd x_2' \frac{|\psi(x_1',x_2')|^2}{|x_i-x_j'|} \, .
\end{align}
We consider the case where the two systems are separated by a distance $d$. 
We are interested in the terms that encode the SN interactions between the two mesoscopic objects. Hence, we will neglect the non-linear self-interaction terms that modify the motion of each object separately \cite{Yang2013}.  The terms we are looking for are the cross terms in \eqref{eq:full-SN-potential}
\begin{align}
\begin{split}
       V^\times_\text{SN}=G\,m^2\left(\int  \dd x_1' \dd x_2' \frac{|\psi(x_1',x_2')|^2}{|d + x_2-x_1'|} \Bigg. 
   \Bigg. + \int \dd x_1'\dd x_2' \frac{|\psi(x_1',x_2')|^2}{|d + x_2'- x_1|} \right) \, . 
\end{split}
\end{align}
 We now suppose that the separation of the two systems $d$ is greater than the size of each object and that the state is well localized with respect to separation so that the denominator can be expanded in a Taylor series, yielding
\begin{align}
\begin{split}
    \int  \dd x_1' \dd x_2' \frac{|\psi(x_1',x_2')|^2}{|d + x_2-x_1'|}= \frac{1}{d}\Bigg(1 + \frac{\langle x_1\rangle- x_2}{d} + \frac{(\langle x_1\rangle- x_2)^2}{2 d^2} + \cdots \Bigg) \, ,
\end{split}
\end{align}
and
\begin{align}
\begin{split}
    \int  \dd x_1' \dd x_2' \frac{|\psi(x_1',x_2')|^2}{|d + x_2'-x_1|}= \frac{1}{d}\Bigg(1 + \frac{ x_1- \langle x_2 \rangle}{d} + \frac{(x_1- \langle x_2 \rangle)^2}{2 d^2} + \cdots \Bigg) \, .
\end{split}
\end{align}
Therefore, combining these terms and neglecting cubic and constant contributions, we arrive at
\begin{align}
\begin{split}
    \hat{V}_\text{SN} = \frac{G m^2}{d^2} (\hat{x}_1 - \hat{x}_2)  + \frac{G m^2}{2 d^3} \left[ \left(\hat{x}_1-\langle\hat{x}_2\rangle\right)^2+\left(\hat{x}_2-\langle\hat{x}_1\rangle\right)^2\right].
\end{split}
\end{align}
This is the same mutual SN interaction potential derived in \cite{Yang2013}. We notice immediately that this Hamiltonian does not act on the joint Hilbert space of the two harmonic oscillators. Hence, product states remain product states throughout the evolution generated by \eqref{eq:SN-interaction}. 
\subsection{Dynamics of two coupled oscillators in the Schrödinger-Newton model: initially vanishing first moments}
We now couple two equal masses trapped in harmonic potentials with identical frequency through the Schrödinger-Newton potential \eqref{eq:SN-interaction}. The full Hamiltonian will be 
\begin{align}
\label{eq:full-SN-hamiltonian}
    \hat{H}=\frac{p_1^2}{2m}+\frac{1}{2}m\omega x_1^2+\frac{p_2^2}{2m}+\frac{1}{2}m\omega x_2^2- \frac{G m^2}{d^2} (\hat{x}_1 - \hat{x}_2)  - \frac{G m^2}{2d^3} \left[ \left(\hat{x}_1-\langle\hat{x}_2\rangle\right)^2+\left(\hat{x}_2-\langle\hat{x}_1\rangle\right)^2\right].
\end{align}
We can elucidate the dynamics under this Hamiltonian if we solve for $\langle x_k(t) \rangle$, with $k\in\{1,2\}$, and plug the solutions back into the potential. To that end, we have to solve the coupled ordinary differential equations (ODEs)
\begin{align}
   &\frac{\dd }{\dd t} \langle x_1\rangle = \frac{\langle p_1\rangle}{m}, \\&\frac{\dd }{\dd t}\langle p_1\rangle =-\langle x_1 \rangle\left(m \omega^2-C_2\right) - C_2\langle x_{2}\rangle+ C_1,\\
   &\frac{\dd }{\dd t}\langle x_2\rangle = \frac{\langle p_2\rangle}{m}, \\&\frac{\dd }{\dd t}\langle p_2\rangle =-\langle x_2 \rangle\left(m \omega^2-C_2\right) - C_2\langle x_{1}\rangle- C_1,
\end{align}
where $C_1=G m^2/d^2$ and $C_2=G m^2/d^3$. Given vanishing expectation values for $t=0$, the solutions to the above ODEs are given by
\begin{align}
    \langle x_1 \rangle = \frac{C_1}{m\omega_\g^2}\left(1-\cos(\omega_\g t)\right), \quad \langle p_1\rangle = \frac{C_1}{\omega_\g}  \sin(\omega_\g t),\quad \langle x_2 \rangle = -\frac{C_1}{m\omega_\g^2}\left(1-\cos(\omega_\g t)\right)\, \quad \langle p_2\rangle =- \frac{C_1}{\omega_\g}  \sin(\omega_\g t) ,
\end{align}
where $\omega_\g = \sqrt{\omega^2-2G m/d^3}$. Therefore, the full Hamiltonian can be written in the form
\begin{align}
    \hat{H} =\sum_{k=1}^2\left( \frac{\hat{p}_k^2}{2m}+ \frac{1}{2} m \tilde{\omega}_\g^2\hat{x}^2_k-\hat{x}_k J(t)\right),
\end{align}
with $\tilde{\omega}_\g = \sqrt{\omega^2-G m/d^3}$, and where we defined the drive function
\begin{align}
  J(t) =C_1 - \frac{ C_1}{m\omega_\g^2}\left(1-\cos(\omega_gt)\right)\, .
\end{align}
In the semi-classical Schrödinger-Newton model, the two interacting oscillators see each other's gravitational effect as a classical driving force.
We can solve the dynamics by going to the Heisenberg picture, where
\begin{align}
    \frac{d}{dt}a_k^\dagger = i\tilde{\omega}_\g a_k^\dagger - \frac{i}{2\tilde{\omega}_\g}J(t),
\end{align}
which can be integrated to give
\begin{align}
    a_k^\dagger(t)=e^{i\tilde{\omega}_\g t}\left(a_k^\dagger(0)-\frac{i}{\sqrt{2 \tilde{\omega}_\g}} \int_0^t e^{-i\tilde{\omega}_\g t'}J(t')\,\text{d}t'\right) \, .
\end{align}
From here, we can see that the variance of the position operator of each oscillator is vanishing and constant in time, namely
\begin{align}
    \Delta \hat{x}_1(t)=\Delta \hat{x}_2(t)=\sqrt{\frac{\hbar}{2 m \omega}}, \quad\forall\, t\in[0,\infty).
\end{align}
Therefore, there is no transfer of squeezing and the states cannot be swapped through the SN interaction. The story is different if the initial states are such that the first moments do not vanish at $t=0$.
\subsection{Dynamics of two coupled oscillators in the Schrödinger-Newton model: initially non-vanishing first moments and violation of the LOCC bound}
\label{appendix:SN-dynamics}
Let us consider the case in which the initial state is $\ket{\Psi(0)}=\ket{\alpha}\otimes\ket{0}$, with $\alpha\in \mathbb{C}$, i.e., 
\begin{align}
    \langle x_1(0) \rangle = x_0 = \sqrt{\frac{2\hbar}{ m \omega}}\text{Re}(\alpha), \quad \langle p_1(0) \rangle = p_0 = \sqrt{\frac{2}{\hbar m \omega}}\text{Im}(\alpha).
\end{align}
Let us further approximate the Hamiltonian \eqref{eq:full-SN-hamiltonian}, so that
\begin{align}
        \hat{H}=\frac{p_1^2}{2m}+\frac{1}{2}m\omega x_1^2+\frac{p_2^2}{2m}+\frac{1}{2}m\omega x_2^2+   \frac{G m^2}{d^3} \left( \hat{x}_1\langle\hat{x}_2\rangle+\hat{x}_2\langle\hat{x}_1\rangle\right),
\end{align}
where we have dropped a constant term, and all linear terms that are not proportional to the first moments -- these amount
to a redefinition of the oscillator frequency $\omega$ proportional to $G m/ d^2$, and can safely be ignored. In this case, the solutions to the ODEs are
\begin{align}
\label{eq:SN-solutions-coherent}
    &\langle x_1\rangle= \frac{1}{2} x_0\left[ \cos\left(\omega_\g^+t\right) +\cos \left(\omega_\g^-t\right)\right]+\frac{p_0}{2m}\left[ \frac{\sin\left(\omega_\g^+ t \right)}{ \,\omega_\g^+}+\frac{\sin\left(\omega_\g^- t\right)}{\,\omega_\g^-}\right], \\
    &\langle p_1\rangle= \frac{1}{2} p_0\left[ \cos\left(\omega_\g^+t\right) +\cos \left(\omega_\g^-t\right)\right]+\frac{1}{2} x_0\left[ m \,\omega_\g^+\sin\left(\omega_\g^+ t \right)+m \,\omega_\g^-\sin\left(\omega_\g^- t \right)\right],\\
    &\langle x_2\rangle= \frac{1}{2} x_0\left[ \cos\left(\omega_\g^+t\right) -\cos \left(\omega_\g^-t\right)\right]+\frac{p_0}{2 m} \left[ \frac{\sin\left(\omega_\g^+ t \right)}{\,\omega_\g^+}-\frac{\sin\left(\omega_\g^- t\right)}{\,\omega_\g^-}\right], \\
    &\langle p_2\rangle= \frac{1}{2} p_0\left[ \cos\left(\omega_\g^+t\right) -\cos \left(\omega_\g^-t\right)\right]+\frac{1}{2} x_0\left[ m \,\omega_\g^+\sin\left(\omega_\g^+ t \right)-m \,\omega_\g^-\sin\left(\omega_\g^- t \right)\right].
\end{align}
where $\omega_\g^\pm=\omega  \sqrt{1\pm\frac{\gamma_\g}{\omega}}$, while the second moments do not evolve in time, and $\Delta \hat{x}_1(t)=\Delta \hat{x}_2(t)=\sqrt{\frac{\hbar}{2 m \omega}}$ and $\Delta \hat{p}_1(t)=\Delta \hat{p}_2(t)=\sqrt{\frac{\hbar m \omega}{2 }}$. This tells us that the state of the system at any time $t$ is in a coherent state. Assuming $\gamma_\g \ll \omega$, which is equivalent to the RWA approximation in the quantum gravitational picture, we can expand the frequency $\omega_\g^\pm\simeq \omega \left(1 \pm \frac{\gamma_\g}{2 \omega}\right)$, and, as discussed in \cite{Gollapudi2024}, one can read the state from the solutions \eqref{eq:SN-solutions-coherent}. Alternatively, by computing the fidelity, we can compare the dynamics obtained through the SN evolution to the quantum gravitational dynamics \eqref{eq:gravity-unitary}. This is represented in \figref{fig:SN-locc-violation}. From here, we see that the SN model cannot be distinguished from the quantum gravitational dynamics when the RWA is in place. Further, by explicitly violating the LOCC bound, we corroborate the main claim made in \cite{Gollapudi2024}, i.e., that the dynamics generated by the Schrödinger-Newton model cannot be classified as LOCC.

\section{Proof of the teleportation bound}
\label{appendix:bound-proof}
In this section, we show the proof of the bound \eqref{eq:ourlocc-bound} following \cite{Chiribella2013}. We want to find the best classical process approximating the transformation $\rho(\alpha)\mapsto \rho'(\alpha)$, where $\rho(\alpha)$ is a state on the system $A$ picked from the ensemble $\{\rho(\alpha),p(\alpha)\}$ and $\rho'(\alpha)$ is an output target state on the system $A'$. We consider a process where both $\rho(\alpha)$ and $\rho'(\alpha)$ are pure states -- but in principle, we could take $\rho(\alpha)$ to be mixed.  By ``best", as we articulate in the main text, we mean the classical process that maximizes the fidelity measure
\begin{align}
    F_{c\ell} =\sup_{\mathcal{E}}\sum_\alpha p(\alpha)\text{Tr}\left[\mathcal{E}(\rho(\alpha))\,\rho'(\alpha)\right],
\end{align}
where $\mathcal{E}:A\rightarrow A'$ is a measure-and-prepare channel. In the reference \cite{Chiribella2013}, the authors show that the bound can be computed using
\begin{align}
\label{eq:CFT}
    F_{c\ell} = \left\|\left(\id\otimes\tau^{-\frac{1}{2}}\right)\sigma\left(\id\otimes\tau^{-\frac{1}{2}}\right)\right\|_\times, 
\end{align}
where
\begin{align}
    \sigma=\sum_\alpha p(\alpha) \rho'(\alpha)\otimes\rho(\alpha), \quad \tau=\sum_\alpha p(\alpha) \rho(\alpha)\, ,
\end{align}
and where $\|A\|_\times=\sup_{\|\psi\|=\|\phi\|=1}\langle\phi|\langle\psi| A|\phi\rangle|\psi\rangle$. We are interested in finding the classical fidelity threshold for the process $|\alpha\rangle\mapsto |g(t)\,\alpha\rangle$, where $g(t)=\sin(\gamma_\g t)$. We start by noting that the average input state $\tau$ is in the Fock basis given by
\begin{align}
    \tau = \frac{\lambda}{1+\lambda}\sum_{n=0}^\infty\left(\frac{1}{1+\lambda}\right)^n |n\lr n| \, ,
\end{align}
while the state $\sigma$ in \eqref{eq:CFT} is given by
\begin{align}
    \sigma=\frac{\lambda}{\pi} \int \dd^2\alpha\, e^{-\lambda|\alpha|^2}|g(t)\alpha\lr g(t)\alpha|\otimes|\alpha\lr\alpha|\, .
\end{align}
Thus, the operator $A_\tau= \left(\id\otimes\tau^{-\frac{1}{2}}\right)\sigma\left(\id\otimes\tau^{-\frac{1}{2}}\right)$ that enters the cross norm in \eqref{eq:CFT} evaluates to
\begin{align}
    A_\tau=\chi\int \frac{\dd^2\alpha}{\pi} \,|g(t)\alpha\lr g(t)\alpha|\otimes|\sqrt{\chi}\, \alpha\lr\sqrt{\chi}\,\alpha|\, ,
\end{align}
where $\chi=1+\lambda$. Using the beamsplitter transformation defined by 
\begin{align}
    B_\theta=e^{\theta\left(b_1 b_2^\dagger - b_1^\dagger b_2\right)}\,,\quad\theta=\tan^{-1}\left(\frac{g(t)}{\sqrt{\chi}}\right),
\end{align} 
and changing the variables to $\alpha\rightarrow\alpha\sqrt{\chi+g(t)^2}$, the integral can be rewritten as
\begin{align}
    A_\tau=\frac{1+\lambda}{1+\lambda+g(t)^2}\int\frac{\dd^2\alpha}{\pi} B_\theta^\dagger(|\alpha\lr\alpha|\otimes|0\lr0|)B_\theta .
\end{align}
For $g:\mathbb{R}\rightarrow [0,1]$, the steps above are well defined. Since coherent states form an overcomplete set, namely $\int \dd^2 \alpha |\alpha\lr\alpha|=\id$, the integral evaluates to the identity, hence 
\begin{align}
    A_\tau=\left(\frac{1+\lambda}{1+\lambda+g(t)^2}\right) B_\theta^\dagger\left(\id\otimes|0\lr0|\right)B_\theta.
\end{align}
Computing the norm of this operator, we see that
\begin{align}
    \|A_\tau\|_\infty=\|A_\tau\|_\times=\frac{1+\lambda}{1+\lambda+g(t)^2},
\end{align}
thus we arrive at the bound \eqref{eq:ourlocc-bound}.

\end{document}